\documentclass[epj]{webofc}
\usepackage[varg]{txfonts}   
\usepackage{graphicx,color} 
\usepackage{bm}       
\usepackage{amsmath}  
\usepackage{amssymb}
\woctitle{21st International Conference on Few-Body Problems in Physics}
\begin{document}
\title{van der Waals and Casimir-Polder interactions between neutrons}
\author{James F. Babb\inst{1}\fnsep\thanks{\email{jbabb@cfa.harvard.edu}} \and
        Mahir S. Hussein\inst{1,2,3,4}\fnsep\thanks{\email{hussein@if.usp.br}} 
}

\institute{ITAMP, Harvard-Smithsonian Center for Astrophysics, MS 14, 60 Garden St., Cambridge, MA 02138
\and
           Instituto de Estudos Avan\c{c}ados, Universidade de S\~{a}o Paulo C. P.
72012, 05508-970 S\~{a}o Paulo-SP, Brazil
\and
           Instituto de F\'{\i}sica,
Universidade de S\~{a}o Paulo, C. P. 66318, 05314-970 S\~{a}o Paulo,
Brazil
\and
Departamento de F\'{i}sica, Instituto Tecnol\'{o}gico de Aeron\'{a}utica, CTA, S\~{a}o Jos\'{e} dos Campos, S.P., Brazil
          }

\abstract{
We investigate the van der Waals interaction between neutrons using the theory of Casimir and Polder, wherein
the potential for asymptotically large separations falls off as the inverse seventh power,
and compare it to the similar interaction between a neutron and a proton,
for which the asymptotic interaction falls off as the inverse fourth power.
Modifications of the formalism to extend the validity to smaller separations using
dynamic electric and magnetic dipole polarizability data are discussed.
}
\maketitle
\section{Introduction}
\label{intro}
The long-range interaction between a hydrogen atom and an electron 
is $-\alpha_H/2R^4$, where $R$ is the separation distance
and $\alpha_H$ is the electric dipole polarizability of the atom.
We use atomic units throughout.
The potential arises from the charge-induced dipole interaction.
The long-range interaction between a neutron and a nucleus arises
in the same way. It is -$\alpha_n Z^2/ 2R^4$, where $Z$ is the nuclear charge
and $\alpha_n$ is the neutron electric dipole polarizability. 
A physical picture of the atomic polarizability is the response
of the electron in the presence of an applied static electric field; a similar
picture for the neutron might be the response of the pion cloud to a static electric field~\cite{DreWal08,GriMcGPhi12}.

In brief, Thaler~\cite{Tha59} and others considered the effect of
the potential on neutron scattering from heavy nuclei.
Experiments on the transmission of neutrons
through lead used the potential above to obtain a value for $\alpha_n$~\cite{SchRieHar91}.
A number of experiments were carried out, essentially dependent
in some way on using this potential  to measure $\alpha_n$,  and these are
reviewed in Refs.~\cite{WisLevSch98,Sch05}.

As expected with the above physical picture,  the effects of external photons can also
be related to the neutron dipole polarizabilities and much recent work is
focused on using Compton scattering from the neutron bound in the deuteron to extract $\alpha_n$ and the magnetic
dipole polarizability $\beta_n$~\cite{Sch05,GriMcGPhi12}.
The Compton scattering amplitudes are
related to the dynamic polarizabilities $\alpha_n(\omega)$ and $\beta_n(\omega)$,
which describe the response of the neutron to the photon of frequency $\omega$,
and are identical to the static polarizability
values at zero frequency~\cite{Sch05}, \textit{viz.} $\alpha_n(0) = \alpha_n$, $\beta_n(0) = \beta_n$.
Theoretical methods using chiral effective field theory ($\chi$EFT) yielded values of the dynamic
polarizabilities~\cite{GriHem02,HilGriHem04,GriMcGPhi12}. 

We are using dynamic polarizabilities from $\chi$EFT
to investigate the long-range interactions between two neutrons
and between a neutron and a proton~\cite{BabHus15x}.
In this short report, we describe the direction of the project.

\section{Casimir-Polder potential}
\label{CP}
The long-range interaction between two neutrons is given formally for asymptotically
large  separations by the Casimir-Polder potential~\cite{FeiSuc70}, where $c=137$ is the speed of light,
\begin{equation}
\label{nn-infty}
V^\infty_{nn}(r) =  - (c/4 \pi)[23(\alpha_n \alpha_n + \beta_n\beta_n)
 -14(\alpha_n\beta_n)] R^{-7}+
  \mathcal{O}(R^{-9}) , \quad R\sim\infty .
\end{equation}
For neutrons, we take $\alpha_{n}=(11.8 \pm 1.1) \times 10^{-4}\; \textrm{fm}^3$ 
and $\beta_n$ is (3.7 $\pm 1.2) \times 10^{-4}\; \textrm{fm}^3$~\cite{OliAgaAms14}. 
[Note that Eq.~(\ref{nn-infty}) is the asymptotic limit of a more general
equation involving dynamic polarizabilities (see, for example, Ref.~\cite{Bab10}) that describes the neutron-neutron interaction for all distances
at separations $r$ sufficiently large that the interaction is purely electromagnetic.]

An early analysis considering parts of
Eq.~(\ref{nn-infty}) and the general equation and its effect on the neutron scattering length was carried out in 1973 
by Arnold~\cite{Arn73}.
At the time of his analysis, $\beta_n$ was unknown and  the accepted value of $\alpha_n$ was twice today's value.
In particular, the $\chi$EFT predicts a dynamic polarizability, $\alpha_n(\omega)$,  which contains a reference to the excited state of the neutron, and this will have an influence on the nn interaction. This hitherto not considered effect is also being investigated.

The long-range interaction between a proton and a neutron is notably different from
Eq.~(\ref{nn-infty}), 
\begin{equation}
\label{pn-infty}
V^\infty_{pn}(R) = - (1/2) \alpha_{n}R^{-4} + (1/4 \pi cM_p)(11\alpha_n + 5\beta_n) R^{-5}  + \mathcal{O}(R^{-7}) , \quad R\sim\infty ,
\end{equation}
where $M_p$ is the proton mass~\cite{BerTar76}.
[Similarly to Eq.~(\ref{nn-infty}), Eq.~(\ref{pn-infty}) is the asymptotic limit of a general equation involving dynamic polarizabilities~\cite{Bab10} that describes
the proton-neutron interaction at all distances sufficiently large that the interaction
is purely electromagnetic.]

A detailed exploration of Eqs.~(\ref{nn-infty}) and (\ref{pn-infty}) and their more general
forms using dynamic polarizabilities for applications to neutron-neutron 
and neutron-proton scattering and for neutron-wall interactions,
is in progress~\cite{BabHus15x}.

\begin{acknowledgement}
The Institute for Theoretical Atomic, Molecular,
and Optical Physics (ITAMP) is supported in part by a grant from the NSF to the Smithsonian Institution and Harvard University.
MSH is also partly supported by the Brazilian funding agencies CAPES/ITA, CNPq, FAPESP, INCT-IQ/MCT, and CEPID-CEPOF.
\end{acknowledgement}

%
%
%

\begin{thebibliography}{14}

\bibitem{DreWal08}
D.~Drechsel, T.~Walcher, Rev. Mod. Phys. \textbf{80}, 731 (2008)

\bibitem{GriMcGPhi12}
H.W. Grie{\ss}hammer, J.A. McGovern, D.R. Phillips, G.~Feldman, Prog. Part. Nucl.
  Phys. \textbf{67}, 841 (2012), arXiv: 1203.6834

\bibitem{Tha59}
R.M. Thaler, Phys. Rev. \textbf{114}, 827 (1959)

\bibitem{SchRieHar91}
J.~Schmiedmayer, P.~Riehs, J.A. Harvey, N.W. Hill, Phys. Rev. Lett.
  \textbf{66}, 1015 (1991)

\bibitem{WisLevSch98}
F.~Wissmann, M.I. Levchuk, M.~Schumacher, Eur. Phys. J. A \textbf{1}, 193
  (1998)

\bibitem{Sch05}
M.~Schumacher, Prog. Part. Nucl. Phys. \textbf{55}, 567 (2005)

\bibitem{GriHem02}
H.W. Grie{\ss}hammer, T.R. Hemmert, Phys. Rev. C \textbf{65}, 045207 (2002)

\bibitem{HilGriHem04}
R.P. Hildebrandt, H.W. Grie{\ss}hammer, T.R. Hemmert, B.~Pasquini, Eur. Phys.
  J. A \textbf{20}, 293 (2004)

\bibitem{BabHus15x}
J.F. {Babb}, M.S. Hussein, in preparation

\bibitem{FeiSuc70}
G.~Feinberg, J.~Sucher, Phys. Rev. A \textbf{2}, 2395 (1970)

\bibitem{OliAgaAms14}
K.A. Olive et~al., Chin. Phys. C
  \textbf{38}, 090001 (2014)

\bibitem{Bab10}
J.F. Babb, \emph{Adv. At. Molec. Opt. Phys.} (Academic Press, San Diego, 2010),
  Vol.~59, p.~1

\bibitem{Arn73}
L.G. Arnold, Phys. Lett. B \textbf{44}, 401 (1973)

\bibitem{BerTar76}
J.~Bernab\'eu, R.~Tarrach, Ann. Phys. (N.Y.) \textbf{102}, 323 (1976)

\end{thebibliography}


\end{document}